\documentclass[prd,nofootinbib,amsfonts]{revtex4}
\usepackage{graphicx,bm,amsmath,color}
\usepackage[latin1]{inputenc}

\newcommand{\be}{\begin{equation}}
\newcommand{\ee}{\end{equation}}
\newcommand{\beq}{\begin{eqnarray}}
\newcommand{\eeq}{\end{eqnarray}}

\begin{document}

\title{Relativistic kinematics beyond Special Relativity}

\author{J.M. Carmona}
\affiliation{Departamento de F\'{\i}sica Te\'orica,
Universidad de Zaragoza, Zaragoza 50009, Spain}
\author{J.L. Cort\'es}
\affiliation{Departamento de F\'{\i}sica Te\'orica,
Universidad de Zaragoza, Zaragoza 50009, Spain}
\author{F. Mercati}
\email{jcarmona@unizar.es, cortes@unizar.es, flavio.mercati@gmail.com}
\affiliation{Departamento de F\'{\i}sica Te\'orica,
Universidad de Zaragoza, Zaragoza 50009, Spain}

\begin{abstract}
In the context of departures from Special Relativity written as a momentum power expansion in the inverse of an ultraviolet energy scale $M$, we derive the constraints that the relativity principle imposes between coefficients of a deformed (but rotational invariant) composition law, dispersion relation, and transformation laws, at first order in the power expansion.
In particular, we find that, at that order, the consistency of a modification of the energy-momentum composition law fixes the modification in the dispersion relation.
We therefore obtain the most generic modification of Special Relativity which is rotational invariant and that preserves the relativity principle at leading order in $1/M$.
\end{abstract}

\maketitle

\section{Introduction}

Corrections to Special Relativity (SR) coming from quantum gravity effects have been predicted and theoretically explored by string theory and quantum gravity research since more than one decade~\cite{qugra}. Experimental tests have also been carried out in search of residual effects at low energies of these high-energy violations of SR~\cite{Mattingly:2005re,Coleman:1998ti}, putting strong constraints to Lorentz violating coefficients in the Standard Model Extension (SME)~\cite{Colladay:1998fq}, a generalization of the Standard Model that allows for violations of Lorentz and CPT symmetry in the framework of a local effective field theory (EFT). The SME, inspired by what happens in the context of string theory, provides a dynamical scenario which assumes that the Lorentz invariance violation (LIV) arises from spontaneous symmetry breaking in a more fundamental theory with Lorentz covariance. This implies that microcausality and the usual conservation laws of energy and momentum are expected to hold in the low energy effective theory. However, since the vacuum breaks Lorentz invariance, this EFT is formulated in a certain system of reference, and the relativistic principle (equality of the dynamical laws derived from this EFT for different inertial observers) does no longer hold.
There are many experiments and observations which are sensitive to the existence of a preferred reference frame and that imply very strong constraints on the LIV~\cite{Kostelecky:2008ts}.

Doubly Special Relativity (DSR) emerged~\cite{dsr} as an alternative way to consider violations of Lorentz invariance. Lorentz symmetry is \emph{deformed} instead of \emph{broken}, in a way that it is still possible to formulate observer-independent laws of physics.
In this case it is much more difficult to find observable effects of the deviations from SR.
The characteristic deformation of DSR is encoded in an energy-momentum dispersion relation which depends on a high-energy (or short-distance) scale which is invariant in the same sense that the speed of light is a relativistic invariant, together with new (deformed) Lorentz transformations which preserve the form of the dispersion relation. In particular, since the presence of a new energy scale in the modified dispersion relation does not necessarily require to deform the rotations, DSR usually considers a deformation only in the transformation laws under boosts, which, owing to their noncompact character, offer less stringent bounds than the usual rotational invariance.\footnote{This is an heuristic argument that is explicitly seen in the constraints one gets for rotational and nonrotational invariant parameters in the SME, see e.g.~\cite{Kostelecky:2008ts}. Up to our knowledge there are no such type of studies in the case of a deformed symmetry, but considering a rotational invariant deformation is a common practice in DSR theories and will prove to be an important algebraic simplification in the analysis presented here.}

Then, the generators of the deformed Lorentz transformations still satisfy the ordinary Lorentz algebra, but are represented non-linearly in momentum space~\cite{AmelinoCamelia:2002an}. This non-linearity is a source of complication in the multi-particle sectors of DSR theories, since one can no longer define the total momentum of two particles as the sum of the individual momenta if it has to transform under boosts as the momentum of a single particle. This makes necessary the introduction of a deformed composition law which will depend on the new scale of the theory, and evidences that the DSR framework goes beyond the EFT paradigm, where usual composition laws apply.

The fact that DSR cannot be embeded in the SME makes it difficult to provide it with a dynamical formulation. Initially, only kinematic considerations in momentum space guided the explorations of DSR effects in quantum gravity phenomenology. In fact the space-time structure underlying the theory was suspected from the start to be rather non-trivial, for example by containing some fundamental noncommutativity~\cite{KowalskiGlikman:2002jr}. More recently, it became clear that nonlocal issues were an important ingredient of DSR models~\cite{nonlocality}. This is another fact reflecting that the DSR proposal goes beyond the EFT framework. Then, in Ref.~\cite{relativelocality} it was introduced the concept of `relative locality' in terms of a non-trivial geometry of momentum space. A spacetime with relative locality seems a natural candidate for a DSR spacetime, but the proposal in Ref.~\cite{relativelocality} was set forth in a rather general way, without implementing a relativity principle.
In fact not every geometry of momentum space is compatible with the relativity principle. An attempt in this direction was presented in Ref.~\cite{locandrelpple}, where it was remarked that an additive conservation law describes an interaction which is local in spacetime, and non-linear corrections to this law cause the locality property to be lost for a general observer. This means that DSR cannot be local in the same sense as SR, and in fact the description of interactions that was given in Ref.~\cite{relativelocality} under the relative locality notion can be used for a DSR theory which is implemented in terms of the so-called auxiliary variables~\cite{Judes:2002bw}, which are non-linear mappings between a physical momentum (transforming according to deformed Lorentz boosts) and an auxiliary momentum which transforms linearly~\cite{locandrelpple}.

However, the use of auxiliary variables is not enough to describe a generic DSR theory, as we will later clarify in the present paper. This makes the connection between DSR and the possible geometry of relative-locality momentum spaces more involved. Since the latter primarily results in modifications of the on-shell relation and of the law of momentum conservation~\cite{relativelocality}, the problem can be more generally formulated as the connection between a theory which, while including a new high-energy scale, implements a relativistic principle (with deformed Lorentz transformations) and the possible modifications of both the dispersion relation and the composition law when the theory is such that it reduces to SR at energies much lower than the new ultraviolet scale. This was the main viewpoint adopted in Ref.~\cite{AmelinoCamelia:2011yi}, where a `golden-rule' for the `DSR-compatibility' of a given geometry of momentum space (in fact, for the compatibility between the implementation of a relativity principle and a given modified dispersion relation and composition law) was derived. This golden-rule was found by considering two physical processes that cannot occur in a theory beyond SR endowed with a relativity principle: photons cannot decay into electron-positron pairs (since this reaction is forbidden in SR, it would imply the existence of a threshold depending on the ultraviolet scale, which is self-contradictory because the energy of the photon can be tuned above or below this observer-invariant threshold with an appropriate boost) and it must always be possible for a photon of any energy to produce electron-positron pairs in interactions with some sufficiently high-energy photons (if it were not so, this would imply again an energy threshold for the switch-off of this reaction, and then the same argument as before applies). However, Ref.~\cite{AmelinoCamelia:2011yi} expressed doubts whether the obtained golden-rule to be satisfied by a modified dispersion relation and a composition law to be DSR-compatible was not only necessary, but also a sufficient condition.

The main result of the present manuscript is a derivation of the constraints that the principle of relativity imposes on the dispersion relation and the composition law in a theory beyond SR. As in Ref.~\cite{AmelinoCamelia:2011yi}, we will work in a scenario of rotational invariant (polynomial) modifications to SR to the leading order in $M$, the scale of new physics. We will consider an implementation of deformed boost transformations such that the ordinary Lorentz algebra is still satisfied. This will allow us to obtain simple relations between the coefficients which parameterize the leading deviations to SR, to which presumably quantum gravity phenomenology will be most sensitive~\cite{dsrfacts}. In doing so, we will re-derive the golden-rule of Ref.~\cite{AmelinoCamelia:2011yi} in a completely different way and will establish that it is not a sufficient condition for the DSR-compatibility of a modified dispersion relation and composition law. The relations we will provide allow one to identify the most generic set of deviations from SR that are still compatible with the relativity principle, at first order in $1/M$.

The outline of the paper is as follows. In Section II we will give the general forms of the modified dispersion relation and composition law for a two-particle system both in the case $1+1$ dimensions, which we will treat separately because of its simplicity, and in the more relevant $3+1$ case. Then in Section III we will examine the constraints that the relativity principle imposes on them, deriving the set of relations that must exist between the coefficients in the dispersion relation and in the composition law to be compatible with a relativity principle. We will also check these relations in specific examples.

There are some points in common between this work and Ref.~\cite{asr}, where a systematic approach to SR compatible with a relativistic principle was presented also at leading order of Planckian effects, referring to that framework as Asymptotic Special Relativity (ASR). That previous work tried to be more general in the sense of including arbitrary functions of energy or momentum instead of simple polynomials in the corrections to SR, but in fact was more restrictive than the present analysis because it only considered the definition of auxiliary (energy and momentum) variables as a way to go beyond SR. That this is not sufficiently general will be clear in Section IV, which therefore establishes that a criticism of DSR that is often heard (that it is just SR rewritten in another set of variables) is unfounded.

Finally, in Section V we will argument how the present work can be generalized beyond the two-particle case and beyond the $M^{-1}$ order and give some concluding remarks.

\section{Modified dispersion relation and composition law}

We will consider a departure from SR which can be expanded in powers of momenta and the inverse of an ultraviolet scale $M$. If one relates these departures with a quantum spacetime structure associated with quantum gravity fluctuations it is natural to identify the ultraviolet scale $M$ with the Planck mass but more general cases could be considered.
We will assume that all the energies are much smaller than the scale $M$ so that the dominant effect of the corrections to SR kinematics are on the first order terms in the $1/M$ expansion.  We will also restrict all the discussion to the case where there are no departures from rotational symmetry.

We will consider the kinematic analysis of reactions with no more than two particles in the initial or final state. As we will comment at the end of the work, there does not seem to be any obstruction to extend the present analysis to reactions involving more than two particles in the initial or final state and to higher orders terms in the $1/M$ expansion.

There are two ingredients in the generalized kinematics: a modification of the energy-momentum relation of a particle (modified dispersion relation) and a modification of the composition law of two momenta (and the associated modification of the energy-momentum conservation law).

\subsection{$1+1$ dimensional case}

Let us start by considering the simpler case of a generalized kinematics in $1+1$ dimensions.
In order to make easier a comparison with a rotational invariant $3+1$ dimensional generalized kinematics, we will impose in the present case invariance under the parity transformation $p_0 \to p_0$, $p_1 \to - p_1$.
With this restriction, the general form for the composition of two momenta $p$, $q$ is
\be
\left[p\oplus q\right]_0 \,=\, p_0 + q_0 + \frac{\beta_1}{M} p_0 q_0 + \frac{\beta_2}{M} p_1 q_1 {\hskip 1cm}   \left[p \oplus q\right]_1 \,=\, p_1 + q_1 + \frac{\gamma_1}{M} p_0 q_1 + \frac{\gamma_2}{M} p_1 q_0
\label{cl1+1}
\ee
where $\beta_1$, $\beta_2$, $\gamma_1$, $\gamma_2$ are dimensionless coefficients and we are implementing the condition
\be
p \oplus q|_{q=0} \,=\, p {\hskip 1cm}  p \oplus q|_{p=0} \,=\, q
\label{cl0}
\ee
on the composition law.

The general form for the dispersion relation of a particle is given by
\be
C(p) \,=\, p_0^2 - p_1^2 + \frac{\alpha_1}{M} p_0^3 + \frac{\alpha_2}{M} p_0 p_1^2 \,=\, \mu^2\,.
\label{dr1+1}
\ee
The generalization of the SR kinematics is parameterized by the six dimensionless coefficients $\beta_1$, $\beta_2$, $\gamma_1$, $\gamma_2$, $\alpha_1$, $\alpha_2$. We want to determine what are the conditions that these coefficients have to satisfy in order to have a generalized kinematics compatible with the relativity principle (absence of a preferred reference frame).

\subsection{$3+1$ dimensional case}

In $3+1$-dimensions the general form of the composition law of two momenta compatible with rotational invariance takes the form
\be
\left[p\oplus q\right]_0 \,=\, p_0 + q_0 + \frac{\beta_1}{M} \, p_0 q_0 + \frac{\beta_2}{M} \, \vec{p}\cdot\vec{q} {\hskip 1cm}   \left[p \oplus q\right]_i \,=\, p_i + q_i + \frac{\gamma_1}{M} \, p_0 q_i + \frac{\gamma_2}{M} \, p_i q_0
+ \frac{\gamma_3}{M} \, \epsilon_{ijk} p_j q_k
\label{cl3+1}
\ee
and the particle dispersion relation is given by
\be
C(p) \,=\, p_0^2 - \vec{p}^{\,2} + \frac{\alpha_1}{M} \, p_0^3 + \frac{\alpha_2}{M} \, p_0 \vec{p}^{\,2} \,=\, \mu^2 \,.
\label{dr3+1}
\ee
All the difference with the 1+1 dimensional case is that one has an additional (seventh) dimensionless coefficient ($\gamma_3$) in the composition law.
Note that this term would be absent in a parity invariant generalized kinematics in $3+1$ dimensions: without it, $[p \oplus q]_i$ would go to $-[p \oplus q]_i$ under the transformation $p_0 \to p_0$, $\vec p \to -\vec p$, $q_0 \to q_0$, $\vec q \to -\vec q$.
However, we will maintain this term in the $3+1$ dimensional case for a more general discussion.

\section{Relativity principle}

Let us see how a relativity principle can be consistently implemented when one has a generalized composition law of momenta and a generalized dispersion relation.

\subsection{$1+1$ dimensional case}

In the $1+1$ dimensional case all one needs is the transformation under a boost (parameterized by $\xi_1$) of a two particle system with momenta $p$, $q$. The main ingredient in the implementation of the relativity principle is that the transformation in general does not act separately on the momenta of each particle. After the transformation the momenta of the particles will be $T^L_q(p)$, $T^R_p(q)$ with subindex $q$ on $T^L$ indicating that the transformed momentum of the particle that had a momentum $p$ can depend on the momentum of the other particle, and similarly for the subindex $p$ in $T^R$. The upper indexes $L$, $R$ indicate the possibility to have a different transformation laws (noncommutativity) for the two momenta.

The condition (\ref{cl0}) on the composition law allows us to derive the transformation under a boost of a one particle system $p \to T(p)$ from the transformation law of a two particle system
\be
T(p) \,=\, T^L_0(p) \,=\, T^R_0(p) \,,
\ee
where the equality comes from the assumption that a one-particle system is equivalent to a two-particle system in which one of the particles has zero momentum.
The general form of the boost transformation of a one particle system at order $1/M$ can be expressed in terms of three dimensionless parameters $\lambda_1$, $\lambda_2$, $\lambda_3$ as
\begin{eqnarray}
\left[T(p)\right]_0 &=& p_0 + p_1 \xi_1 + \frac{\lambda_1}{M} \, p_0 p_1 \xi_1 \nonumber \\
\left[T(p)\right]_1 &=& p_1 + p_0 \xi_1 + \frac{\lambda_2}{M} \, p_0^2 \xi_1 + \frac{\lambda_1 + 2\lambda_2 + 3\lambda_3}{M} p_1^2 \xi_1
\label{Tlambda}
\end{eqnarray}
where a choice of coefficients has been made in order to make easier the comparison of the $1+1$ and $3+1$ dimensional cases.

The invariance of the dispersion relation (\ref{dr1+1}) under a boost transformation requires that
\be
C(T(p)) \,=\, C(p) \,.
\label{drinv}
\ee
This fixes the dimensionless coefficients $\alpha_1$, $\alpha_2$ in the modified dispersion relation in terms of the three parameters $\lambda_1$, $\lambda_2$, $\lambda_3$ of the boost transformation~(\ref{Tlambda}):
\be
\alpha_1 \,=\, -2 \,(\lambda_1 + \lambda_2 + 2\lambda_3) {\hskip 1cm}
\alpha_2 \,=\, 2 \, (\lambda_1 + 2\lambda_2 + 3\lambda_3)\,.
\label{alpha(lambda)}
\ee

The general form of the boost transformation of a two particle system at order $1/M$ is given by
\be
T^L_q(p) \,=\, T(p) + {\bar T}^L_q(p) {\hskip 1cm} T^R_p(q) \,=\, T(q) + {\bar T}^R_p(q)
\ee
with
\be
\left[{\bar T}^L_q(p)\right]_0 \,=\, \frac{\eta_1^L}{M} \, q_0 p_1 \xi_1 + \frac{\sigma_1^L}{M} \, q_1 p_0 \xi_1 {\hskip 1cm} \left[{\bar T}^L_q(p)\right]_1 \,=\, \frac{\sigma_2^L}{M} \, q_0 p_0 \xi_1 + \frac{\sigma_3^L}{M} \, q_1 p_1 \xi_1 \,,
\label{TbarL1+1}
\ee
\be
\left[{\bar T}^R_p(q)\right]_0 \,=\, \frac{\eta_1^R}{M} \, p_0 q_1 \xi_1 + \frac{\sigma_1^R}{M} \, p_1 q_0 \xi_1 {\hskip 1cm} \left[{\bar T}^R_p(q)\right]_1 \,=\, \frac{\sigma_2^R}{M} \, p_0 q_0 \xi_1 + \frac{\sigma_3^R}{M} \, p_1 q_1 \xi_1 \,.
\label{TbarR1+1}
\ee
But the invariance of the dispersion relation of each of the particles under a boost transformation requires that
\be
C\left({\bar T}^L_q(p)\right) \,=\, C\left({\bar T}^R_p(q)\right) \,=\, 0 \,.
\label{drinv2}
\ee
This implies that
\be
\sigma_1^L \,=\, \sigma_3^L \,=\, \sigma_1^R \,=\, \sigma_3^R \,=\, 0 {\hskip 1cm}
\sigma_2^L \,=\, \eta_1^L {\hskip 1cm} \sigma_2^R \,=\, \eta_1^R
\ee
and then one has
\begin{eqnarray}
\left[{\bar T}^L_q(p)\right]_0 &=& \frac{\eta_1^L}{M} \, q_0 p_1 \xi_1 {\hskip 1cm}
\left[{\bar T}^L_q(p)\right]_1 \,=\, \frac{\eta_1^L}{M} \, q_0 p_0 \xi_1 \nonumber \\
\left[{\bar T}^R_p(q)\right]_0 &=& \frac{\eta_1^R}{M} \, p_0 q_1 \xi_1 {\hskip 1cm}
\left[{\bar T}^R_p(q)\right]_1 \,=\, \frac{\eta_1^R}{M} \, p_0 q_0 \xi_1
\label{Tbar1+1}
\end{eqnarray}
so that finally one has only two new dimensionless parameters $\eta_1^L$, $\eta_1^R$ in the boost transformation of a two particle system.

The last step in the implementation of the relativity principle is the requirement that
\be
T(p\oplus q) \,=\, T^L_q(p) \oplus T^R_p(q) \,.
\label{clinv}
\ee
This guarantees the invariance under boosts of the energy-momentum conservation in the decay of one particle into two. It is a straightforward algebraic problem to determine the relations that Eq.~(\ref{clinv}) implies between the parameters of the boost transformations and the dimensionless coefficients in the composition law of momenta:
\begin{align}
\beta_1 &= 2 \,(\lambda_1 + \lambda_2 + 2\lambda_3) &
\beta_2 &= -2 \lambda_3 - \eta_1^L - \eta_1^R \nonumber \\
\gamma_1 &= \lambda_1 + 2\lambda_2 + 2\lambda_3 - \eta_1^L &
\gamma_2 &= \lambda_1 + 2\lambda_2 + 2\lambda_3 - \eta_1^R \,.
\label{clT1+1}
\end{align}

It is clear then that the boost transformation of the one- and two-particle system, determined
by the $\lambda$ and $\eta$ coefficients,  fix completely the momentum composition law ($\gamma$ and $\beta$ coefficients). DSR models, for which the deformed boost transformation of the one particle system is well defined, usually contain an ambiguity in the energy-momentum conservation law for particle processes. We see that, in the context of the $1/M$ expansion that we are dealing with, this ambiguity corresponds to the deformation in the boost of the two particle system ($\eta_1^L$ and $\eta_1^R$ coefficients) that one may consider. Once the transformation laws for the one particle and two particle systems are given, both the dispersion relation and the composition law are fixed.

On the other hand, one could consider the composition law, which gives the conservation laws in an interacting system, as more fundamental than the transformation laws from a physical point of view. Then we see that, for a given composition law ($\beta$ and $\gamma$ coefficients) there is a one-parameter family of deformed transformation laws that can implement a relativity principle. Defining $\lambda=(\lambda_1+\lambda_2)/2$, then the coefficients that define these transformation laws are
\begin{alignat}{2}
\lambda_1 &= \frac{3\beta_1}{4}+\frac{\beta_2}{2}-\frac{\gamma_1}{2}-\frac{\gamma_2}{2}+\lambda \quad \quad \quad \quad
\lambda_2 &= -\frac{3\beta_1}{4}-\frac{\beta_2}{2}+\frac{\gamma_1}{2}+\frac{\gamma_2}{2}+\lambda \quad \quad \quad \quad
\lambda_3 &= \frac{\beta_1}{4}-\lambda \nonumber \\
\eta_1^L &= -\frac{\beta_1}{4}-\frac{\beta_2}{2}-\frac{\gamma_1}{2}+\frac{\gamma_2}{2}+\lambda \quad \quad \quad \quad
\eta_1^R &= -\frac{\beta_1}{4}-\frac{\beta_2}{2}+\frac{\gamma_1}{2}-\frac{\gamma_2}{2}+\lambda \quad \quad \quad \quad &
\end{alignat}

When one combines these relations with the expressions (\ref{alpha(lambda)}) of the coefficients in the modified dispersion relation in terms of the parameters of the boost transformations one gets
\be
\alpha_1 \,=\, - \beta_1 {\hskip 2cm} \alpha_2 \,=\, \gamma_1 + \gamma_2 - \beta_2 \,.
\label{RP}
\ee
The relativity principle implemented through (\ref{drinv}), (\ref{drinv2}), (\ref{clinv}) fixes the modification in the dispersion relation in terms of the modification in the composition law. This remark has never been made in the DSR literature before and is one of the main results of our work.

A consequence of the choice of a modified dispersion relation consistent with a modified composition law is the relation
\be
\alpha_1 + \alpha_2 + \beta_1 + \beta_2 - \gamma_1 - \gamma_2 \,=\, 0 \,.
\label{goldenrule}
\ee
This is just the `golden-rule' derived in Ref.~\cite{AmelinoCamelia:2011yi} as a consequence of the incompatibility of the relativity principle with the existence of an energy threshold for the decay of one particle into two (``no-photon-decay-switch-on constraint'') and with the existence of an energy threshold for the energy of one particle in the production of two particles by two particles (``no-pair-production-switch-off constraint''). In fact in Ref.~\cite{AmelinoCamelia:2011yi}  it is conjectured, based on a few examples, that the golden rule~(\ref{goldenrule}) is not only a necessary but also a sufficient condition for the compatibility of a departure from SR with the relativity principle.

The systematic derivation of the modification in the dispersion relation induced by a modification in the composition law and the compatibility with the relativity principle gives an alternative derivation of the golden rule independent of considerations of thresholds in particular reactions. It also shows that the golden rule (\ref{goldenrule}) \underline{is not a sufficient condition} for a relativistic kinematics but just a combination of the two relations (\ref{RP}) that fix the modified dispersion relation compatible with a given composition law.

\subsection{Some examples}

We can check explicitly that the examples considered in Ref.~\cite{AmelinoCamelia:2011yi} to test the validity of the golden rule and its possible sufficiency are particular cases of the implementation of the relativity principle as discussed in this section. In the first example, know in DSR literature as `DSR1'~\cite{Bruno:2001mw}, one has a boost transformation\footnote{We are identifying $1/M$ with the invariant length $\ell$; this is just a convention since all the relations are invariant under a  simultaneous rescaling of all the dimensionless coefficients, together with a corresponding rescaling of $M$.}
\be
\left[T(p)\right]_0 \,=\, p_0 + p_1 \xi_1 {\hskip 1cm} \left[T(p)\right]_1 \,=\, p_1 + p_0 \xi_1 +
\frac{1}{M} p_0^2 \xi_1 + \frac{1}{2 M} p_1^2 \xi_1 \,.
\ee
This corresponds in our notation to
\be
\lambda_1 \,=\, 0 {\hskip 1cm} \lambda_2 \,=\, 1 {\hskip 1cm} \lambda_3 \,=\, - \frac{1}{2} \,. \label{lambdaDSR1}
\ee
The modified dispersion relation corresponds to
\be
C(p) \,=\, p_0^2 - p_1^2 + \frac{1}{M} p_0 p_1^2\,,
\ee
that is,
\be
\alpha_1 \,=\, 0 {\hskip 1cm} \alpha_2 \,=\, 1 \,. \label{alphaDSR1}
\ee
The choice (\ref{lambdaDSR1})  and (\ref{alphaDSR1})  is compatible with the relations (\ref{alpha(lambda)}). For the composition law one has
\be
\left[p\oplus q\right]_0 \,=\, p_0 + q_0 + \frac{1}{M} p_1 q_1 {\hskip 1cm}
\left[p\oplus q\right]_1 \,=\, p_1 + q_1 + \frac{1}{M} p_0 q_1 + \frac{1}{M} p_1 q_0
\ee
so that
\be
\beta_1 \,=\, 0 {\hskip 1cm} \beta_2 \,=\, 1 {\hskip 1cm} \gamma_1 \,=\, 1 {\hskip 1cm} \gamma_2 \,=\, 1 \,.
\ee
One can easily check that the relations (\ref{RP}) are satisfied and then one has a modified dispersion relation compatible with the modification in the composition law. One can also check that the relations (\ref{clT1+1}) between dimensionless coefficients in the composition law and parameters in the boost transformation are satisfied with
\be
\eta_1^L \,=\, \eta_1^R \,=\,0 \,.
\ee
Then one has in this case
\be
{\bar T}^L_q(p) \,=\, {\bar T}^R_p(q) \,=\, 0 {\hskip 1cm} T(p\oplus q) \,=\, T(p)\oplus T(q)
\ee
and the boost transformation of the two particle system reduces to an independent boost transformation on each of the particles.

The second example considered in Ref.~\cite{AmelinoCamelia:2011yi} corresponds to
\be
\left[T(p)\right]_0 \,=\, p_0 + p_1 \xi_1 - \frac{1}{M} p_0 p_1 \xi_1 {\hskip 1cm} \left[T(p)\right]_1 \,=\, p_1 + p_0 \xi_1 +
\frac{1}{M} p_0^2 \xi_1 + \frac{1}{M} p_1^2 \xi_1 \,,
\ee
and then
\be
\lambda_1 \,=\, -1 {\hskip 1cm} \lambda_2 \,=\, 1 {\hskip 1cm} \lambda_3 \,=\, 0 \,.
\ee
The modified dispersion relation is
\be
C(p) \,=\, p_0^2 - p_1^2 + \frac{2}{M} p_0 p_1^2\,,{\hskip 1cm} \alpha_1 \,=\, 0 {\hskip 1cm} \alpha_2 \,=\, 2\,,
\ee
and the modified composition composition law in this example is
\be
\left[p\oplus q\right]_0 \,=\, p_0 + q_0  {\hskip 1cm}
\left[p\oplus q\right]_1 \,=\, p_1 + q_1 + \frac{1}{M} p_0 q_1 + \frac{1}{M} p_1 q_0\,,
\ee
so that
\be
\beta_1 \,=\, 0 {\hskip 1cm} \beta_2 \,=\, 0 {\hskip 1cm} \gamma_1 \,=\, 1 {\hskip 1cm} \gamma_2 \,=\, 1\,.
\ee
Also in this example the relation (\ref{RP}) is satisfied. The relations (\ref{clT1+1}) lead also in this case to
\be
\eta_1^L \,=\, \eta_1^R \,=\, 0 {\hskip 1cm} {\bar T}^L_q(p) \,=\, {\bar T}^R_p(q) \,=\, 0 {\hskip 1cm} T(p\oplus q) \,=\, T(p)\oplus T(q)
\ee
and then to a trivial boost transformation for the two particle system as in the first example.

The other two examples considered in Ref.~\cite{AmelinoCamelia:2011yi} are inspired by studies of the $\kappa$-Poincar\'e Hopf algebra.
The third example has an additive composition law of energies as in the previous example but the composition of momentum is noncommutative:
\be
\left[p\oplus q\right]_0 \,=\, p_0 + q_0  {\hskip 1cm}
\left[p\oplus q\right]_1 \,=\, p_1 + q_1 + \frac{1}{M} p_0 q_1\,,
\ee
so that
\be
\beta_1 \,=\, 0 {\hskip 1cm} \beta_2 \,=\, 0 {\hskip 1cm} \gamma_1 \,=\, 1 {\hskip 1cm} \gamma_2 \,=\, 0 \,.
\ee
The boost transformation acting on a particle is given by
\be
\left[T(p)\right]_0 \,=\, p_0 + p_1 \xi_1 {\hskip 1cm}
\left[T(p)\right]_1 \,=\, p_1 + p_0 \xi_1 + \frac{1}{M} p_0^2 \xi_1 + \frac{1}{2M} p_1^2 \xi_1\,,
\ee
and then
\be
\lambda_1 \,=\, 0 {\hskip 1cm} \lambda_2 \,=\, 1 {\hskip 1cm} \lambda_3 \,=\, -\frac{1}{2} \,.
\ee
The modified dispersion relation compatible with this boost transformation is
\be
C(p) \,=\, p_0^2 - p_1^2 + \frac{1}{M} p_0 p_1^2 \,, {\hskip 1cm} \alpha_1 \,=\, 0 {\hskip 1cm} \alpha_2 \,=\, 1
\ee
and once more these coefficients of the modified dispersion relation are compatible with the relativity principle constraints (\ref{RP}). From the equations (\ref{clT1+1}) which give the modified composition law in terms of the boost transformations we can see that the transformation of a two particle system is given in this case by
\be
\eta_1^L \,=\, 0 {\hskip 1cm} \eta_1^R \,=\, 1
\ee
and then
\be
T(p\oplus q) \,=\, T(p) \oplus T^R_p(q) {\hskip 1cm} \left[{\bar T}^R_p(q)\right]_0 \,=\, \frac{1}{M} p_0 q_1 \xi_1 {\hskip 1cm}  \left[{\bar T}^R_p(q)\right]_1 \,=\, \frac{1}{M} p_0 q_0 \xi_1\,,
\ee
so that we have a nontrivial transformation of the two particle system. Note that in general one has from (\ref{clT1+1}),
\be
\gamma_1 - \gamma_2 \,=\, \eta_1^L - \eta_1^R
\ee
and then a noncommutativity ($\gamma_1\neq \gamma_2$) in the momentum composition law automatically implies a nontrivial boost transformation of the two particle system.

The last example in Ref.~\cite{AmelinoCamelia:2011yi} corresponds to an unmodified dispersion relation ($\alpha_1=\alpha_2=0$) and an additive composition law of energies. The constraints from the relativity principle (\ref{RP}) require $\gamma_1=-\gamma_2$ so that one has
\be
\left[p\oplus q\right]_0 \,=\, p_0 + q_0 {\hskip 1cm}
\left[p\oplus q\right]_1 \,=\, p_1 + q_1 + \frac{1}{2M} p_0 q_1 - \frac{1}{2M} p_1 q_0 \,.
\ee
The boost transformation of a particle state is given by
\be
\left[T(p)\right]_0 \,=\, p_0 + p_1 \xi_1 + \frac{1}{2M} p_0 p_1 {\hskip 1cm}
\left[T(p)\right]_1 \,=\, p_1 + p_0 \xi_1 + \frac{1}{2M} p_0^2
\ee
so that
\be
\lambda_1 \,=\, \frac{1}{2} {\hskip 1cm} \lambda_2 \,=\, \frac{1}{2} {\hskip 1cm} \lambda_3 \,=\, - \frac{1}{2}
\ee
and
\be
\eta_1^L \,=\, 0 {\hskip 1cm} \eta_1^R \,=\, 1\,,
\ee
which corresponds to the same boost transformation for the two particle system as in the previous example.

The fact that in all four examples the golden rule (\ref{goldenrule}) is satisfied is directly a consequence of the relativity principle present in all the examples. It is very easy to introduce a modification in the dispersion relation or the composition law in each of the four examples such that the golden rule (\ref{goldenrule}) is still satisfied but not the two relations (\ref{RP}) between the coefficients in the dispersion relation and the coefficients in the composition law: in all these cases one would have a modification of the SR kinematics consistent with the golden rule but not with the relativity principle, signaling the presence of a preferred reference frame.

\subsection{$3+1$ dimensional case}

The general form the transformation of one particle at order $1/M$, which now depends on three parameters $\vec{\xi}$,  is in this case
\begin{eqnarray}
\left[T(p)\right]_0 &=& p_0 + (\vec{p} \cdot \vec{\xi}) + \frac{\lambda_1}{M} \, p_0 (\vec{p} \cdot \vec{\xi}) \nonumber \\
\left[T(p)\right]_i &=& p_i + p_0 \xi_i + \frac{\lambda_2}{M} \, p_0^2 \xi_i + \frac{\lambda_3}{M} \, {\vec p}^{\,2} \xi_i + \frac{\lambda_4}{M} \, p_i ({\vec p} \cdot {\vec \xi}) + \frac{\lambda_5}{M} \, p_0 \epsilon_{ijk} p_j \xi_k \,.
\end{eqnarray}
In the case of $3+1$ dimensions one has an additional restriction for $T$ to correspond to a boost transformation. In order to reproduce the Lorentz algebra one has a condition on the composition of two transformations $T^{(1)}$, $T^{(2)}$ with parameters ${\vec \xi}^{\,(1)}$,  ${\vec \xi}^{\,(2)}$:
\begin{eqnarray}
&&\left[T^{(2)}\left(T^{(1)}(p)\right) - T^{(1)}\left(T^{(2)}(p)\right)\right]_0 \,=\, 0 \nonumber \\
&&\left[T^{(2)}\left(T^{(1)}(p)\right) - T^{(1)}\left(T^{(2)}(p)\right)\right]_i \,=\,  ({\vec p} \cdot {\vec \xi}^{\,(1)}) \, {\xi^{(2)}}_i  -   ({\vec p} \cdot {\vec \xi}^{\,(2)}) \, {\xi^{(1)}}_i \,.
\label{LaT}
\end{eqnarray}
This requires that
\be
\lambda_5 \,=\, 0 {\hskip 1cm} \lambda_4 \,=\, \lambda_1 + 2\lambda_2 + 2\lambda_3
\ee
and the boost transformation of a one particle system is then
\begin{eqnarray}
\left[T(p)\right]_0 &=& p_0 + (\vec{p} \cdot \vec{\xi}) + \frac{\lambda_1}{M} \, p_0 (\vec{p} \cdot \vec{\xi}) \nonumber \\
\left[T(p)\right]_i &=& p_i + p_0 \xi_i + \frac{\lambda_2}{M} \, p_0^2 \xi_i + \frac{\lambda_3}{M} \, {\vec p}^{\,2} \xi_i + \frac{(\lambda_1 + 2\lambda_2 + 2\lambda_3)}{M} \, p_i ({\vec p} \cdot {\vec \xi}) \,.
\label{T3+1}
\end{eqnarray}
The invariance of the modified dispersion relation (\ref{dr3+1}) under boosts fixes the coefficients $\alpha_1$, $\alpha_2$
\be
\alpha_1 \,=\, -2 \,(\lambda_1 + \lambda_2 + 2\lambda_3) {\hskip 1cm}
\alpha_2 \,=\, 2 \, (\lambda_1 + 2\lambda_2 + 3\lambda_3)
\label{alpha(lambda)3+1}
\ee
which are exactly the expressions found in the $1+1$ dimensional case.\footnote{This is the reason for the choice of coefficients in $1+1$ dimensions.}

For a transformation of a two particle system we have the same general structure of the $1+1$ dimensional case with
\begin{eqnarray}
\left[{\bar T}^L_q(p)\right]_0 &=& \frac{\eta_1^L}{M} \, q_0 ({\vec p} \cdot {\vec \xi}) + \frac{\sigma_1^L}{M} \, p_0 ({\vec q} \cdot {\vec \xi}) + \frac{\eta_2^L}{M} \, ({\vec p} \wedge {\vec q}) \cdot {\vec \xi} \nonumber \\
\left[{\bar T}^L_q(p)\right]_i &=&  \frac{\eta_3^L}{M} \, q_i ({\vec p} \cdot {\vec \xi}) + \frac{\sigma_2^L}{M} \, p_i ({\vec q} \cdot {\vec \xi}) + \frac{\eta_4^L}{M} \, q_0 \epsilon_{ijk} p_j \xi_k + \frac{\sigma_3^L}{M} \, ({\vec p} \cdot {\vec q}) \xi_i + \frac{\sigma_4^L}{M} \, p_0 \epsilon_{ijk} q_j \xi_k + \frac{\sigma_5^L}{M} \, p_0 q_0 \xi_i
\end{eqnarray}
and similar expresions for ${\bar T}^R_p(q)$. The invariance of the dispersion relation when we replace $p$ by $T^L_q(p)$ requires that $C({\bar T}^L_q(p))=0$. This implies that
\be
\sigma_1^L \,=\, \sigma_2^L \,=\,  0 {\hskip 1cm} \sigma_3^L \,=\, - \eta_3^L {\hskip 1cm} \sigma_4^L \,=\,  \eta_2^L {\hskip 1cm} \sigma_5^L \,=\, \eta_1^L \,,
\ee
so that we have
\begin{eqnarray}
\left[{\bar T}^L_q(p)\right]_0 &=& \frac{\eta_1^L}{M} \, q_0 ({\vec p} \cdot {\vec \xi}) + \frac{\eta_2^L}{M} \, ({\vec p} \wedge {\vec q}) \cdot{\vec \xi} \nonumber \\
\left[{\bar T}^L_q(p)\right]_i &=& \frac{\eta_1^L}{M} \, p_0 q_0 \xi_i + \frac{\eta_2^L}{M} \,  p_0 \epsilon_{ijk} q_j \xi_k + \frac{\eta_3^L}{M} \, \left[q_i ({\vec p} \cdot {\vec \xi}) - ({\vec p} \cdot {\vec q}) \xi_i\right] + \frac{\eta_4^L}{M} \, q_0 \epsilon_{ijk} p_j \xi_k
\end{eqnarray}
(and a similar expression for ${\bar T}^R_p(q)$) replacing (\ref{Tbar1+1}) in $1+1$ dimensions. We have three new coefficients $\eta^L_2$, $\eta^L_3$, $\eta^L_4$  that multiply terms vanishing identically in $1+1$ dimensions (the 1+1 dimensional limit can be taken by putting all the $2$ and $3$ components of $\vec p$, $\vec q$ and $\vec \xi$ to zero). But there is still another condition to have a relativistic kinematics. The transformation $p \to T^L_q(p)$ has to be consistent with the Lorentz algebra. This last condition implies that
\be
\eta_3^L \,=\, \eta_1^L {\hskip 1cm} \eta_4^L \,=\, - \eta_2^L\,,
\ee
so that finally we have just two  additional coefficients ($\eta_2^L$, $\eta_2^R$)  in the $3+1$  boost transformations of a two particle system
\begin{eqnarray}
\left[{\bar T}^L_q(p)\right]_0 &=& \frac{\eta_1^L}{M} \, q_0 ({\vec p} \cdot {\vec \xi}) + \frac{\eta_2^L}{M} \, ({\vec p} \wedge {\vec q}) \cdot{\vec \xi}
{\hskip 1cm} \left[{\bar T}^R_p(q)\right]_0 = \frac{\eta_1^R}{M} \, p_0 ({\vec q} \cdot {\vec \xi}) - \frac{\eta_2^R}{M} \, ({\vec p} \wedge {\vec q}) \cdot {\vec \xi}
\nonumber \\
\left[{\bar T}^L_q(p)\right]_i &=& \frac{\eta_1^L}{M} \, p_0 q_0 \xi_i + \frac{\eta_2^L}{M} \,  \left[p_0 \epsilon_{ijk} q_j \xi_k  - q_0 \epsilon_{ijk} p_j \xi_k\right] +
\frac{\eta_1^L}{M} \, \left[q_i ({\vec p} \cdot {\vec \xi}) - ({\vec p} \cdot {\vec q}) \xi_i\right] \nonumber \\
\left[{\bar T}^R_p(q)\right]_i &=& \frac{\eta_1^R}{M} \, p_0 q_0 \xi_i - \frac{\eta_2^R}{M} \,  \left[p_0 \epsilon_{ijk} q_j \xi_k  - q_0 \epsilon_{ijk} p_j \xi_k\right] +
\frac{\eta_1^R}{M} \, \left[p_i ({\vec q} \cdot {\vec \xi}) - ({\vec p} \cdot {\vec q}) \xi_i\right]\,.
\label{Tbar3+1}
\end{eqnarray}

The last step in the implementation of the relativity principle is to enforce the invariance under boosts of the energy-momentum conservation in the decay of one particle into two
\be
T(p\oplus q) \,=\, T^L_q(p) \oplus T^R_p(q)\,.
\ee
A straightforward algebra (just more tedious than in the $1+1$ dimensional case) allows to determine the five dimensionless coefficients $\beta_1$, $\beta_2$, $\gamma_1$, $\gamma_2$, $\gamma_3$ of the $3+1$ dimensional composition laws (\ref{cl3+1}) in terms of the parameters $\lambda_1$, $\lambda_2$, $\lambda_3$, $\eta_1^L$, $\eta_1^R$, $\eta_2^L$, $\eta_2^R$  of the boost transformations (\ref{T3+1}), (\ref{Tbar3+1}). The results are
\begin{alignat}{3}
\beta_1 &= 2 \,(\lambda_1 + \lambda_2 + 2\lambda_3) \quad\quad &
\beta_2 &= -2 \lambda_3 - \eta_1^L - \eta_1^R \quad\quad & \nonumber \\
\gamma_1 &= \lambda_1 + 2\lambda_2 + 2\lambda_3 - \eta_1^L \quad\quad &
\gamma_2 &= \lambda_1 + 2\lambda_2 + 2\lambda_3 - \eta_1^R \quad\quad & \gamma_3 = \eta_2^L - \eta_2^R
\label{clT3+1}
\end{alignat}
which are the same results obtained in the $1+1$ dimensional case for $\beta_1$, $\beta_2$, $\gamma_1$, $\gamma_2$ in terms of  $\lambda_1$, $\lambda_2$, $\lambda_3$, $\eta_1^L$, $\eta_1^R$ and an additional expression for the coefficient $\gamma_3$ (which does not appear in $1+1$ dimensions) as a function of the new ($3+1$ dimensional) parameters in the boost transformations $\eta_2^L$, $\eta_2^R$. Since the expressions for the coefficients $\alpha_1$, $\alpha_2$ in the dispersion relation as a function of the parameters $\lambda_1$, $\lambda_2$, $\lambda_3$ in the one particle boost transformations are also the same as in $1+1$ dimensions one concludes that the derivation of the modified dispersion relation from a modified conservation law in $1+1$ dimensions is also valid in $3+1$ dimensions without any change. The only additional ingredient when going to $3+1$ dimensions is that there is a new additional source of noncommutativity in the momentum composition law ($\gamma_3\neq 0$) together with the one already present in $1+1$ dimensions ($\gamma_1-\gamma_2\neq 0$). As in the case of $1+1$ dimensions one sees that a noncommutative momentum composition law requires a nontrivial boost transformation of the two particle system (${\bar T}^L_q(p) \neq 0$ or ${\bar T}^R_p(q) \neq 0$).
In this case, there is a two-parameter family of transformation laws for the one particle and two particle systems which are compatible with a given composition law in a relativistic theory.

\section{Generalized kinematics versus SR kinematics}

Lacking a complete consistent dynamical framework incorporating the previous discussion of the generalized relativistic kinematics, a question arises whether the generalized kinematics reduces to SR in a nontrivial (e.g. nonlinear) choice of energy-momentum variables. In fact it is not clear whether one has a freedom for such a choice of variables or if the variables used in the modified expression of the composition law have a physical dynamical content. Letting aside this crucial question one can formally analyze how the composition law of momenta varies when one makes a general change of energy-momentum variables at order $1/M$.

In $1+1$ dimensions one can introduce new variables $P_0$, $P_1$ through
\be
p_0 \,=\, P_0 + \frac{\delta_1}{M} P_0^2 + \frac{\delta_2}{M} P_1^2 {\hskip 1cm} p_1 \,=\, P_1 + \frac{\delta_3}{M} P_0 P_1 \,.
\label{aux}
\ee
At the same time, the previous change of variables allows us to give an interpretation to the symbols $\left[P\oplus Q\right]_0$ and $\left[P\oplus Q\right]_1$. Since $(p\oplus q)$ is also a momentum, applying Eq.~(\ref{aux}) to it gives
\begin{eqnarray}
\left[p\oplus q\right]_0 &=& \left[P\oplus Q\right]_0 + \frac{\delta_1}{M} \left[P\oplus Q\right]_0^2 + \frac{\delta_2}{M} \left[P\oplus Q\right]_1^2 \,, \nonumber \\
\left[p\oplus q\right]_1 &=& \left[P\oplus Q\right]_1 + \frac{\delta_3}{M} \left[P\oplus Q\right]_0 \left[P\oplus Q\right]_1 \,.
\end{eqnarray}
Then the composition law (\ref{cl1+1}), when rewritten in terms of the new variables, takes the form
\begin{eqnarray}
\left[P\oplus Q\right]_0 &=& P_0 + Q_0 + \frac{(\beta_1-2\delta_1)}{M} P_0Q_0 + \frac{(\beta_2-2\delta_2)}{M} P_1Q_1 \nonumber \\
\left[P \oplus Q\right]_1 &=& P_1 + Q_1 + \frac{(\gamma_1-\delta_3)}{M} P_0Q_1 + \frac{(\gamma_2-\delta_3)}{M} P_1Q_0 \,.
\end{eqnarray}

In the case of a commutative momentum composition law ($\gamma_1=\gamma_2$) it is possible to choose new variables by
\be
\delta_1 \,=\, \beta_1/2 {\hskip 1cm} \delta_2 \,=\, \beta_2/2 {\hskip 1cm} \delta_3 \,=\, \gamma_1 \,=\, \gamma_2
\label{delta}
\ee
so that the composition law of momenta is additive in the new variables.

For the dispersion relation~(\ref{dr1+1}), Eq.~(\ref{aux}) gives
\begin{equation}
\mu^2 =   P_0^2 - P_1^2
+ \frac{2\delta_1+\alpha_1}{M} P_0^3 + \frac{2(\delta_2-\delta_3)+\alpha_2}{M}  P_0 P_1^2\,.
\end{equation}
For the new variables with an additive composition law, relations~(\ref{delta}) allow us to write
\begin{equation}
\mu^2 =  P_0^2 - P_1^2
+ \frac{\beta_1+\alpha_1}{M} P_0^3 + \frac{\beta_2-\gamma_1-\gamma_2+\alpha_2}{M}  P_0 P_1^2\,,
\end{equation}
and then the dispersion relation takes the standard unmodified SR form as a consequence of the relativity principle Eq.~(\ref{RP}).

Therefore, for a commutative momentum composition law,
it is possible to change variables to an ``auxiliary energy-momentum'' which behaves just as the SR energy-momentum. In this case,
the possibility of a generalization of SR kinematics is based on the physical dynamical meaning of the choice of energy-momentum variables. However, if one has a noncommutative momentum composition ($\gamma_1\neq \gamma_2$), then the physical content of the generalized kinematics is manifest.

The previous results can be extended to the $3+1$ dimensional case. We have
 \be
p_0 \,=\, P_0 + \frac{\delta_1}{M} P_0^2 + \frac{\delta_2}{M} {\vec P}^{\, 2} {\hskip 1cm} p_i \,=\, P_i + \frac{\delta_3}{M} P_0 P_i
\ee
and
\begin{eqnarray}
\left[P\oplus Q\right]_0 &=& P_0 + Q_0 + \frac{(\beta_1-2\delta_1)}{M} P_0Q_0 + \frac{(\beta_2-2\delta_2)}{M} ({\vec P} \cdot {\vec Q}) \nonumber \\
\left[P \oplus Q\right]_i &=& P_i + Q_i + \frac{(\gamma_1-\delta_3)}{M} P_0Q_i + \frac{(\gamma_2-\delta_3)}{M} P_iQ_0 + \frac{\gamma_3}{M} \epsilon_{ijk} P_j Q_k\,, \nonumber \\
\mu^2 &=&   P_0^2 - |\vec P|^2
+ \frac{2\delta_1+\alpha_1}{M} P_0^3 + \frac{2(\delta_2-\delta_3)+\alpha_2}{M}  P_0 |\vec{P}|^2\,.
\label{aux3+1}
\end{eqnarray}

In the case of a commutative composition of momenta ($\gamma_1-\gamma_2=\gamma_3=0$) the same choice of new variables as in $1+1$ dimensions~(\ref{delta}),
together with the relativity principle~(\ref{RP}),
leads to unmodified composition laws and dispersion relations so that the physical content of the generalization of the kinematics rests on the noncommutativity of the momentum composition or on the dynamical content of the choice of energy-momentum variables.

The message of this section can be rephrased in the language of the geometry of momentum
space as introduced in Refs. \cite{relativelocality}: we have basically asked when a composition law can be described by a flat connection on a Minkowski momentum space, but in a generalized coordinate set. The generality of the answer is limited by the choice we have made throughout this
paper to work at first order in $M^{-1}$. At this order one is not sensitive
to the curvature of momentum space (which can enter only at order $M^{-2}$).
There is still space for a torsion and nonmetricity of momentum space, which are first-order effects. Our result was, unsurprisingly, that our momentum space is Minkowski if the torsion is zero (in fact the torsion measures the noncommutativity of the composition law and $\gamma_1-\gamma_2=\gamma_3=0$ imply that it is zero).
Since in this case the relativity principle~(\ref{RP}) sends the dispersion relation in Eq.~(\ref{aux3+1}) to
$\mu^2=P_0^2- |\vec{P}|^2$, we conclude that the case of a commutative composition law is compatible with a geometry of momentum space which is Minkowski with the metric connection.

\section{Concluding remarks}

In this work we have studied the constraints that the relativity principle imposes on the dispersion relation and the composition law in a
(rotational invariant) theory beyond SR. In particular, we have shown that the composition law fixes completely the dispersion relation but not the transformation laws, so that in $3+1$ dimensions there is a two-parameter family of transformation laws compatible with a given composition law. On the other hand, giving the transformation laws for the one particle and two particle systems, fixes everything.

It is easy to see that, at the order $1/M$ for which we have derived these results, there is no further freedom in the definition of the transformation laws and the composition laws for a system of more than two particles. The restrictions
\be
p\oplus q \oplus r|_{r=0}=p\oplus q \quad \quad p\oplus q \oplus r|_{q=0}=p\oplus r \quad \quad p\oplus q \oplus r|_{p=0}=q\oplus r
\ee
fix completely the composition law of three momenta at order $1/M$ in terms of the composition law of two momenta:
\begin{align}
[p\oplus q \oplus r]_0 & = p_0+q_0+r_0+\frac{\beta_1}{M}(p_0 q_0+p_0r_0 + q_0r_0)+\frac{\beta_2}{M}(p_1 q_1+p_1r_1 + q_1r_1) \nonumber\\
[p\oplus q \oplus r]_1 & = p_1+q_1+r_1+\frac{\gamma_1}{M}(p_0 q_1+p_0r_1 + q_0r_1)+\frac{\gamma_2}{M}(p_1 q_0+p_1r_0 + q_1r_0)
\end{align}
and also it fixes the implementation of the boost transformation in the three particle system in terms of the boost transformation of the two particle system
\be
\{p,q,r\} \to \{T_{q,r}^{(1)}(p),T_{p,r}^{(2)}(q),T_{p,q}^{(3)}(r)\}
\ee
with
\be
T_{q,r}^{(1)}(p)=T(p)+\bar{T}_{q,r}^{(1)}(p) \quad \quad
T_{p,r}^{(2)}(q)=T(q)+\bar{T}_{p,r}^{(2)}(q) \quad \quad
T_{p,q}^{(3)}(r)=T(r)+\bar{T}_{p,q}^{(3)}(r)
\ee
and
\be
\bar{T}_{q,r}^{(1)}(p)=\bar{T}_q^L(p)+\bar{T}_r^L(p) \quad \quad
\bar{T}_{p,r}^{(2)}(q)=\bar{T}_p^R(q)+\bar{T}_r^L(q) \quad \quad
\bar{T}_{p,q}^{(3)}(r)=\bar{T}_p^R(r)+\bar{T}_q^R(r).
\ee
This result obviously extends to the $3+1$ dimensional case.

In terms of the geometric interpretation of the ``relative locality'' proposal \cite{relativelocality}, the above results are a consequence of the
fact that all the novelty that can enter in three-particle vertices is contained in the possibility of a nonassociativity of the composition law, which encodes the curvature of the connection (its Riemann
tensor). But at first order in $1/M$ there is no distinction between
a flat and a curved connection, and therefore all the properties of three- (or several-) particle
systems are encoded by those of the two-particle system.

One can also try to generalize the discussion of the $1/M$ relativistic kinematics to the next order in the power expansion in $1/M$. The number of coefficients for the composition law, dispersion law and transformation laws gets much higher, complicating the algebra, but there is not any obstruction for the analysis. In this case one has a composition law for two particles and a new composition law for three particles which is no longer fixed by the two particle law. In the geometric picture one would say that the (unspecified) components of the curvature enter in the composition law.

It could be interesting to establish the correspondence between the present framework for relativistic kinematics based on a modification of the momentum composition law and a framework based on symmetry algebras generalizing the Poincar\'e symmetry algebra of SR. The derivation of boost transformations of multiparticle systems and modified energy-momentum conservation laws from a $\kappa$-Poincar\'e Hopf algebra presented in Ref.~\cite{Gubitosi:2011ej} should correspond to a particular case of the framework presented in this paper.

A modification of the momentum composition (and conservation) law, requires a new implementation of translational symmetry and a generalization of the SR notion of spacetime. It is an open question how a physically meaningful spacetime should be introduced in a generalized relativistic kinematics. There are even doubts~\cite{AmelinoCamelia:2012qw} on the possible redundancy of such a notion.

Finally, in the paper we gave some hints on the interpretation of some of our results in terms of the geometry of momentum space. It would be very interesting to translate all  the consistency conditions we found for a  dispersion relation and a composition law of momenta  to be compatible with the relativity principle into conditions on the geometry of momentum space. The first task would be to calculate the relationship that holds, at first order in $1/M$, between our coefficients $\alpha,\beta,\gamma$ and the components of all the geometric tensors that define momentum space: the metric,
the torsion and the nonmetricity (as we already remarked, the curvature is unspecified at that order).
Then one would need to ``mod out'' by diffeomorphisms, or general coordinate transformations.
Hopefully one should be able to end up with a simple characterization of the most generic geometry of momentum space that is compatible with the relativity principle.
This issue as well as the study of new examples of generalized kinematics constructed from the general framework presented in this work is presently under investigation.


\section*{Acknowledgments}
This work is supported by CICYT (grant FPA2009-09638) and DGIID-DGA (grant 2011-E24/2).



\end{document}